\documentclass{JHEP3}

\usepackage{epsfig}
\usepackage{amsmath}
\usepackage{amssymb}

\title{An invariant approach to dynamical fuzzy spaces \\ with a three-index variable}

\preprint{YITP-05-34 \\  hep-th/0506192}

\author{Naoki Sasakura \\ Yukawa Institute for Theoretical Physics, Kyoto University,
Kyoto 606-8502, Japan \\ E-mail: \email{sasakura@yukawa.kyoto-u.ac.jp}}

\abstract{
A dynamical fuzzy space might be described by a three-index variable ${C_{ab}}^c$, 
which determines
the algebraic relations $f_a f_b ={C_{ab}}^c f_c$ among the functions $f_a$ on the fuzzy space. 
A fuzzy analogue of the general coordinate transformation would be given by the general linear transformation 
on $f_a$. 
I study equations for the three-index variable invariant under the general linear transformation, and
show that the solutions can be generally constructed from the invariant tensors of Lie groups.  
As specific examples, I study SO(3) symmetric solutions, 
and discuss the construction of a scalar field theory on a fuzzy two-sphere within this framework.
}

\begin{document}

%%%%%% private definition %%%%%%%
\def\CR{\nonumber \\}
\def\pt{\partial}
\def\be{\begin{equation}}
\def\ee{\end{equation}}
\def\bea{\begin{eqnarray}}
\def\eea{\end{eqnarray}}
\def\eq#1{(\ref{#1})}
\def\la{\langle}
\def\ra{\rangle}
\def\hyp{\hbox{-}}
\def\sixj#1#2#3#4#5#6{\left\{ \begin{matrix} #1&#2&#3 \\ #4&#5&#6 \end{matrix} \right\}}
\def\threej#1#2#3#4#5#6{\left( \begin{matrix} #1&#2&#3 \\ #4&#5&#6 \end{matrix} \right)}
%%%%%%%%%%%%%%%%%%%%%%%%%%%%%%%%%

\section{Introduction and a general proposal}
\label{introduction}
Several thought experiments in semi-classical quantum gravity and string theory 
\cite{Garay}-\cite{Sasakura:1999xp} suggest that the classical notion of space-time in general relativity 
should be replaced with a quantum one in quantum treatment of space-time dynamics.
An interesting candidate is given by the non-commutative geometry 
or the fuzzy space \cite{Connes}-\cite{Landi:1997sh}.  
The central principle of general relativity is the invariance under the general coordinate transformation.
It would be obviously interesting if a fuzzy space can be obtained from formulation invariant 
under a fuzzy analog of the general coordinate transformation.
This kind of formulation would lead to fuzzy general relativity, 
which may describe the gravitational dynamics and 
the evolution of the universe in terms of dynamical fuzzy spaces\footnote{The present author considered 
evolving fuzzy spaces in \cite{Sasakura:2003ke}-\cite{Sasakura:2004yr}.}.

A fuzzy space is characterized by the algebraic relations $f_a f_b={C_{ab}}^c f_c$ among 
the functions $f_a$ on the fuzzy space.
Therefore it would be a natural assumption that a dynamical fuzzy space can be described by
treating the three-index variable ${C_{ab}}^c$ dynamically. 
It will be argued in the following section that
a fuzzy analog of the general coordinate transformation is given by  
the general linear transformation on $f_a$. Thus the central proposal 
of this paper is that dynamical fuzzy spaces are described by equations for ${C_{ab}}^c$
invariant under the general linear transformation.

In the following section, a fuzzy analog of the general coordinate 
transformation will be discussed and the general form of the models will be proposed.  
In Section \ref{thesolutions}, it will be shown that the solutions to the equations 
can be constructed from the invariant tensors of Lie groups. 
In Section \ref{so3case}, the Lie group will be specified to SO(3), and some series of the solutions will be 
explicitly constructed. 
In Section \ref{fuzzytwosphere}, I will discuss the construction of a scalar field theory on a fuzzy two-sphere 
by using the result in the preceding section.
The final section will be devoted to discussions and comments.

\section{A fuzzy general coordinate transformation and the general form of the models}
\label{themodel}
Let me first review the general coordinate transformation in the usual commutative space $R^d$. 
A basis of the continuous functions on $R^d$ can be given by 
\be
\{ 1, x^i, x^ix^j, x^ix^jx^k,\cdots \},
\ee
where $x^i\ (i=1,\cdots,d)$ are the coordinates of $R^d$. Let $f_a$ with an index $a$ denote these independent 
functions in the set.
A continuous function on $R^d$ is given by a linear combination of $f_a$.

Let me consider a general coordinate transformation,
\be
\label{coordtrans}
{x'}^i=f^i(x).
\ee
It is a natural restriction that the coordinate transformation is not singular, and is invertible.
Since the right-hand side is a continuous function, the general coordinate transformation can be
represented by a linear transformation,
\be
\label{transx}
{x'}^i={M_i}^a f_a,
\ee
where ${M_i}^a$ are real.

Let me now consider a fuzzy space with a finite number of independent functions $f_a\  (a=1,\cdots,n)$
on the fuzzy space. This is a fuzzy space corresponding to a compact space in usual continuous theory.   
I assume that all the $f_a$ denote real functions on a fuzzy space, 
and that the variable ${C_{ab}}^c$, which determines the algebraic relations
$f_a f_b ={C_{ab}}^c f_c$, is real. I do not assume the associativity or the commutativity of the 
algebra, so that ${C_{ab}}^c$ has no other constraints.

An important assumption of this paper is to interpret the transformation rule \eq{transx} as a partial appearance 
of a more general linear transformation,
\be
\label{transm}
f'_a={M_a}^b f_b,
\ee
where ${M_a}^b$ can take any real values provided that the matrix ${M_a}^b$ is invertible, which comes 
from the assumed invertibility  of the coordinate transformation \eq{coordtrans}. 
Therefore the fuzzy general coordinate transformation of this paper is given by the GL($n$,$R$) 
transformation on $f_a$.
Under the transformation \eq{transm}, the three-index variable ${C_{ab}}^c$ transforms in the way,
\be
\label{transc}
{{C'_{ab}}^c}={M_a}^{a'}{M_b}^{b'} {C_{a'b'}}^{c'}{ (M^{-1})_{c'}}^c.
\ee
I impose that the equations of motion for ${C_{ab}}^c$ must be invariant under this GL($n$,$R$) transformation.

For convenience, let me introduce a graphical expression. The three-index variable ${C_{ab}}^c$ can be 
graphically represented as in Fig.\,\ref{fig1}.
\EPSFIGURE{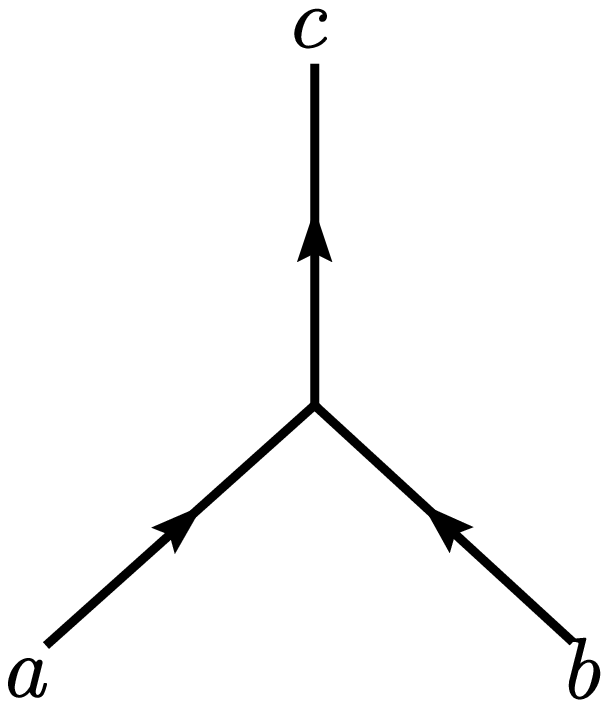, scale=.6}{The graphical representation of ${C_{ab}}^c$. The orientation shows
whether the associated index is a lower or upper one. 
The order of $a$ and $b$ can be also read in the diagram.\label{fig1}}
An example of an invariant equation of motion is given by 
\be
\label{exeq}
{C_{ia}}^j {C_{jb}}^k {C_{kc}}^i+{C_{ai}}^j {C_{cj}}^k {C_{bk}}^i=0,
\ee
which is graphically represented in Fig.\,\ref{fig2}.
It is clear that such invariant equations of motion can be made in infinitely many ways. 

In the next, let me discuss the construction of an action. From the transformation property \eq{transc}, the 
lower and upper indices must be contracted to make an invariant under the GL($n$,$R$) transformation.
Since an action must be invariant and ${C_{ab}}^c$ has more lower indices  
than upper, it is necessary to introduce an additional variable which has more upper indices to 
construct an invariant action. 
A way to achieve this is to define such a variable from the ${C_{ab}}^c$ itself. For example, one can define 
\be
\label{gcrelation}
({C_{ai}}^j{C_{jb}}^i)^{-1},
\ee
which has two upper indices.
\EPSFIGURE{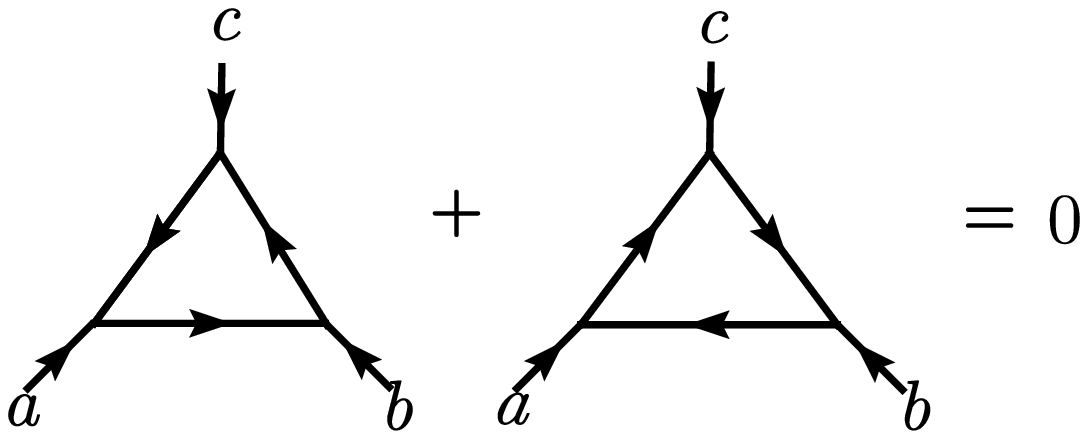,scale=.6}{The graphical representation of the example equation \eq{exeq}. 
The oriented lines connecting vertices represent the contracted indices.
\label{fig2}}
This procedure, however,  will not work when the matrix in the parentheses in \eq{gcrelation}
is not invertible, and will also let an action be a complicated function of ${C_{ab}}^c$.   
Therefore, rather than defining a quantity like \eq{gcrelation} from the beginning,  
I simply introduce a new variable with upper indices 
$g^{ab}$ and assume it be determined from the equations of motion derived from an action. 
I impose its reality and the symmetry of the two indices,
\be
\label{gsym}
g^{ab}=g^{ba}.
\ee
The grahical representation of $g^{ab}$ is given in Fig.\,\ref{fig3}.
\EPSFIGURE{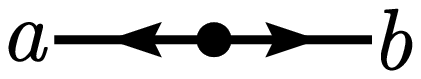,scale=.7}{The graphical representation of the variable $g^{ab}$.
\label{fig3}}

Thus an invariant action is a function of the two dynamical real variables $g^{ab}$ and ${C_{ab}}^c$,
\be
\label{generalaction}
S(g^{ab}, {C_{ab}}^c),
\ee
where all the indices are contracted. 
The graphical representation of an invariant action is 
a closed diagram with oriented lines connecting blobs and three-vertices.

\section{The classical solutions to the equations of motion}
\label{thesolutions}
The quantum mechanical treatment of the model presented in the previous section would be  
obviously very interesting, but in this paper I restrict myself to the classical solutions to 
the equations of motion
derived from the invariant action \eq{generalaction}. Let me suppose that the equations of motion 
for ${C_{ab}}^c$, 
\be
\label{eomgeneralofc}
\frac{\partial S}{\partial {C_{ab}}^c}= 0,
\ee
are satisfied at ${C_{ab}}^c={C^0_{ab}}^c,\ g^{ab}=g_0^{ab}$.  For simplicity, let me assume $g_0^{ab}$ 
is invertible as a matrix.
At the general values of ${C_{ab}}^c$ and $g^{ab}$, 
the invariance of the action under the GL($n$,$R$) transformation implies
\be
\label{wicg}
\frac{\partial S}{\partial g^{ab}} \delta g^{ab} + \frac{\partial S}{\partial {C_{ab}}^c} 
\delta {C_{ab}}^c=0,
\ee
where $\delta g^{ab}$ and $\delta {C_{ab}}^c$ are the infinitesimally small GL($n$,$R$) transformation
of $g^{ab}$ and ${C_{ab}}^c$,
\bea
\label{delgc}
\delta g^{ab}&=&-\delta {M_i}^a g^{ib} - \delta {M_i}^b g^{ai}, \cr
\delta {C_{ab}}^c&=& \delta {M_a}^i {C_{ib}}^c+\delta {M_b}^i {C_{ai}}^c-\delta {M_i}^c {C_{ab}}^i,
\eea
where $\delta {M_i}^a$ can take any  infinitesimally small real values.
At the solution to \eq{eomgeneralofc}, ${C_{ab}}^c={C^0_{ab}}^c,\ g^{ab}=g_0^{ab}$, \eq{wicg} becomes
\be
\delta {M_i}^a g_0^{ib} \left.\frac{\partial S}{\partial g^{ab}}\right|
_{\renewcommand\arraystretch{0.5} \begin{matrix} \scriptstyle g=g_0 \\ \scriptstyle C=C^0 \end{matrix}}=0,
\ee
where I have used \eq{gsym}, \eq{eomgeneralofc} and \eq{delgc}.
Therefore, since $\delta{M_i}^a$ is arbitrary and $g_0^{ab}$ is invertible, 
\be
\left.\frac{\partial S}{\partial g^{ab}}\right|_{\renewcommand\arraystretch{0.5} \begin{matrix} 
\scriptstyle g=g_0 \\ \scriptstyle C=C^0 \end{matrix}}=0
\ee
are satisfied.
This means that the equations of motion for $g^{ab}$ are always
simultaneously satisfied when the equations of motion for ${C_{ab}}^c$ are satisfied. 
Therefore it is enough to consider only the equations of motion \eq{eomgeneralofc} 
to find the classical solutions, provided $g^{ab}$ is invertible.

Now let me consider a Lie group which has a representation of dimension $n$. The representation can be 
either reducible or irreducible. Let me embed the representation into a classical solution: 
The lower index of a classical solution is assumed to be transformed in 
the representation, while the upper one in the inverse representation.
In fact, the following discussions do not depend on whether the representation is real or complex, provided that 
the invariant tensors considered below are real.  
Let $g_0^{ab}$ be a real symmetric rank-two invariant tensor under the inverse representation. 
I assume the tensor $g_0^{ab}$ is invertible as a matrix. 
Let me introduce ${I^\alpha_{ab}}^c\ (\alpha=1,2,\cdots,N)$, 
which span all the real tensors invariant under the same transformation property as ${C^0_{ab}}^c$.
Let me assume ${C^0_{ab}}^c$ is given by a linear combination of these invariant tensors,
\be
\label{caexpression}
{C^0_{ab}}^c= A_\alpha {I^\alpha_{ab}}^c,
\ee
where $A_\alpha\ (\alpha=1,2,\cdots,N)$ are real coefficients.
Since the action is obviously invariant under the transformation of the Lie group, 
the left-hand side of \eq{eomgeneralofc} becomes
a real invariant tensor when $g^{ab}$ and ${C_{ab}}^c$ are substituted with the invariant tensors above.
Therefore, after exchanging the upper and lower indices by using $g_0^{ab}$ and its inverse $g^0_{ab}$,
the left-hand side of \eq{eomgeneralofc} can be expressed as a linear combination of
${I^\alpha_{ab}}^c$,
\be
g^0_{ai} g^0_{bj} g_0^{ck} \left.\frac{\partial S}{\partial {C_{ij}}^k}
\right|_{\renewcommand\arraystretch{0.5} \begin{matrix} \scriptstyle g=g_0 \\ \scriptstyle C=C^0 \end{matrix}}=
 B_\alpha(A_1,A_2,\cdots,A_N){I^\alpha_{ab}}^c,
\ee
where $B_\alpha(A_1,A_2,\cdots,A_N)$ are some functions of $A_1,A_2,\cdots,A_N$, which are
determined from the specific form of the left-hand side of \eq{eomgeneralofc}.
Therefore the equations of motion \eq{eomgeneralofc} are reduced to the following
simultaneous equations for $A_1,A_2,\cdots,A_N$,
\be
\label{simbeq}
B_\alpha(A_1,A_2,\cdots,A_N)=0,\ \  (\alpha=1,\cdots,N).
\ee  
These equations are much easier to solve than \eq{eomgeneralofc}.
Since the numbers of the variables and the equations are the same, the simultaneous equations \eq{simbeq}
will generally have some number of solutions. The solutions are generally complex, 
but real solutions can be actually found in some interesting cases.  
 
\section{Explicit solutions for SO(3)}
\label{so3case}
The Lie algebra of SO(3) is given by
\be
[J_z,J_\pm]=\pm J_\pm,\ \ [J_+,J_-]=2 J_z, 
\ee
and an irreducible representation labeled with an integer spin $j$ can be explicitly given by
\bea
J_z|j,m \rangle&=&m\, |j,m\rangle, \cr
J_\pm |j,m\rangle&=&\sqrt{j(j+1)-m(m\pm 1)}\,|j,m\pm 1\rangle.
\eea
It is convenient to associate the index of the model with a pair $(j,m)$, where $j$ is the spin, 
and $m$ the $J_z$ eigenvalue $(m=-j,-j+1,\cdots,j)$. 
The above representation of SO(3) is generally complex but the invariant tensors can be taken real,
\bea
\label{defofgi}
g_0^{(j_1,m_1)\, (j_2,m_2)}&=& \delta_{j_1\, j_2}\delta_{m_1\,-m_2} (-1)^{m_1}, \cr
I_{(j_1,m_1)\,(j_2,m_2)\,(j_3,m_3)}&=&\left( \begin{array}{ccc} j_1&j_2&j_3 \\ m_1&m_2&m_3 \end{array} \right),
\eea
where the right-hand side in the second line is the $3j$-symbol \cite{Messiah,Varshalovich:1988ye}. 
These invariant tensors are essentially 
unique when the spins of the representations are given. 

Let me consider a representation given by the direct sum of a number of the irreducible representations. 
Here I consider the case that each irreducible representation appears at most once in the direct sum.
Then the ansatz \eq{caexpression} becomes
\be
\label{assofc}
C^0_{(j_1,m_1)\,(j_2,m_2)\,(j_3,m_3)}=A(j_1,j_2,j_3)\,I_{(j_1,m_1)\,(j_2,m_2)\,(j_3,m_3)},
\ee
where the spins $j_i$ must be contained in the representation.
Since the 3$j$-symbol has the cyclic symmetry $1\rightarrow 2\rightarrow 3\rightarrow 1$, 
this is also imposed on $A(j_1,j_2,j_3)$.  It is assumed that $A(j_1,j_2,j_3)$ vanishes
if $j_i$ do not satisfy the triangle inequalities because $I_{(j_1,m_1)\,(j_2,m_2)\,(j_3,m_3)}$
vanishes identically.

The 3$j$-symbol satisfies the identity,
\bea
\label{threethreej}
\sum_{M_i} (-1)^{\sum_{i=1}^3 l_i+M_i}
\left( \begin{array}{ccc} l_1&l_2&j_3 \\ M_1&-M_2&m_3 \end{array} \right) 
\left( \begin{array}{ccc} l_2&l_3&j_1 \\ M_2&-M_3&m_1 \end{array} \right)   
\left( \begin{array}{ccc} l_3&l_1&j_2 \\ M_3&-M_1&m_2 \end{array} \right)  
=\cr
\left( \begin{array}{ccc} j_1&j_2&j_3 \\ m_1&m_2&m_3 \end{array} \right)      
\left\{ \begin{array}{ccc} j_1&j_2&j_3 \\ l_1&l_2&l_3 \end{array} \right\},
\eea
where $\{:::\}$ denotes the $6j$-symbol.
This identity can be used to compute
\be
D^0_{abc}=C^0_{ail}C^0_{bki'}C^0_{cl'k'} g_0^{ii'} g_0^{ll'} g_0^{kk'}
\ee
in Fig.\,\ref{fig4}, where the roman indices abbreviate the pairs $(j,m)$.
The result is 
\bea
\label{evald}
&&D^0_{(j_1,m_1)\, (j_2,m_2)\, (j_3,m_3)} =(-1)^{\sum_{i=1}^3 j_i} 
\left( \begin{array}{ccc} j_1&j_2&j_3 \\ m_1&m_2&m_3 \end{array} \right)\cr
&& \ \ \ \ \ \ \ \ \ \ \ \ \ \times \sum_{l_i} (-1)^{\sum_{i=1}^3 l_i} 
\left\{ \begin{array}{ccc} j_1&j_2&j_3 \\ l_1&l_2&l_3 \end{array} \right\}
A(j_1,l_3,l_2)A(j_2,l_1,l_3)A(j_3,l_2,l_1).
\eea
\EPSFIGURE{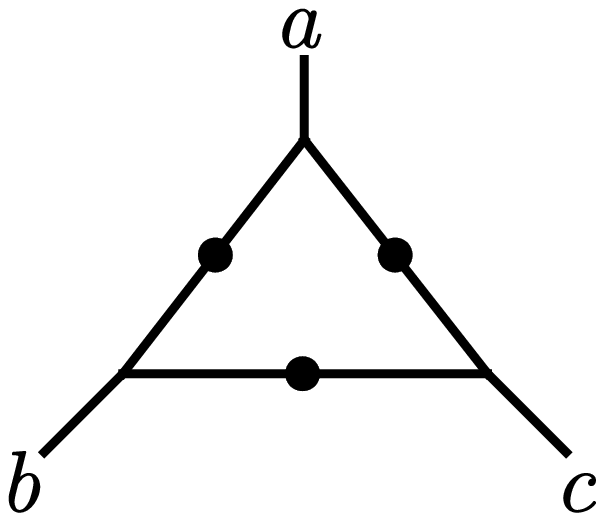,scale=0.7}{The graphical representation of $D^0_{abc}$.  The arrows can be neglected, 
because $C^0_{abc}$ in \eq{assofc} has the cyclic symmetry.
\label{fig4}}
In the above derivation I have also used the property that the 3$j$-symbol in
\eq{defofgi} changes its sign by $(-1)^{\sum_{i=1}^3 j_i}$, when a pair of the rows are interchanged.
Let me consider the equations of motion,
\be
\label{eomcd}
C^0_{abc}-g D^0_{abc}=0,
\ee 
where $g$ is a real coupling constant. The graphical representation is given in Fig.\,\ref{fig5}.
The action which leads to the equations of motion \eq{eomcd} can be constructed in several ways.
This will not be discussed here, because the classical solutions are the main interest in this paper.
From the result \eq{evald}, the equations of motion are reduced to 
\be
\label{eombya}
A(j_1,j_2,j_3)-g
\sum_{l_i} (-1)^{\sum_{i=1}^3 l_i+j_i} 
\left\{ \begin{array}{ccc} j_1&j_2&j_3 \\ l_1&l_2&l_3 \end{array} \right\}
A(j_1,l_3,l_2)A(j_2,l_1,l_3)A(j_3,l_2,l_1)=0.
\ee

A series of the solutions to \eq{eombya} parameterized by a spin parameter $L$ 
can be constructed in the following way. The 6$j$-symbol satisfies the identity,
\bea
\sum_{l_i} (-1)^{\sum_{i=1}^3 l_i} \prod_{i=1}^3 (2l_i+1)
\left\{ \begin{matrix} j_1&j_2&j_3 \\ l_1&l_2&l_3 \end{matrix} \right\}
\left\{ \begin{matrix} j_1&l_3&l_2 \\ b&a_2&a_3 \end{matrix} \right\}
\left\{ \begin{matrix} j_2&l_1&l_3 \\ b&a_3&a_1 \end{matrix} \right\}
\left\{ \begin{matrix} j_3&l_2&l_1 \\ b&a_1&a_2 \end{matrix} \right\} \cr
=(2b+1) (-1)^{b+\sum_{i=1}^3 a_i+j_i} \left\{ \begin{matrix} j_1&j_2&j_3 \\ a_1&a_2&a_3 \end{matrix} \right\}.
\eea
\EPSFIGURE{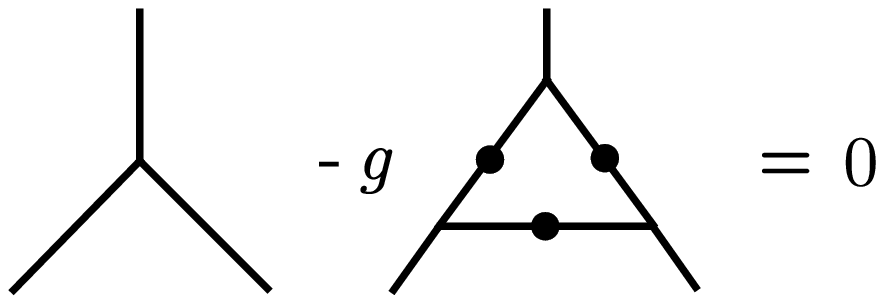,scale=.7}{The equations of motion \eq{eomcd}.
\label{fig5}} 
Therefore, 
\be
\label{seriessol}
A(j_1,j_2,j_3)=\frac{1}{\sqrt{(2L+1)\, g}}\prod_{i=1}^3 \sqrt{2j_i+1}
\left\{ \begin{matrix} j_1&j_2&j_3 \\ L&L&L \end{matrix} \right\}
\ee
is the solution to \eq{eombya} for any given spin $L$. Note that $\forall j_i\leq 2L$ for $A(j_1,j_2,j_3)$ 
to be non-zero, 
because the six arguments of the 6$j$-symbol must be the edge lengths of a tetrahedron.
Therefore the solution \eq{seriessol} can be embedded in a finite-dimensional model
where the index of the model runs over $(j,m)\ (j=0,1,\cdots,2L;\ m=-j,-j+1,\cdots,j)$.

I will now discuss another series of the solutions which will be used in the construction
of a scalar field theory on a fuzzy two-sphere.
The only non-vanishing components of ${C^0_{ab}}^c$ are assumed to be 
\be
\label{coftwosphere}
{C^0_{(1,m_1)\,(L,m_2)}}^{(L,m_3)}=A \threej 1LL{m_1}{m_2}{-m_3} g_0^{(L,-m_3)\, (L,m_3)},
\ee
where $L$ is a given spin and $A$ is a real coefficient, which will be determined from the equations of motion.
The rank-two symmetric tensor $g_0^{ab}$ is as before. The index of the model runs over $(1,m)\ (m=-1,0,1),
\ (L,m)\ (m=-L,-L+1,\cdots,L)$.
The cyclic symmetry for the indices of ${C^0_{ab}}^c$ is not imposed this time, and 
${C^0_{ab}}^c$ are non-zero only when its first index has a spin one and the others a spin $L$.
The equations of motion considered are 
\be
\label{sphereeom}
{C_{ab}}^c-g \, {C_{ai}}^j {C_{ki'}}^{b'} {C_{k'c'}}^{j'} g^{ii'} g_{jj'} g^{kk'} g_{bb'} g^{cc'}=0,
\ee
where $g$ is the coupling constant and $g_{ab}$ is the inverse of $g^{ab}$. Note that the invertibility
of $g^{ab}$ must be assumed from the beginning in defining this model.
The graphical representation is given in Fig.\,\ref{fig6}.
\EPSFIGURE{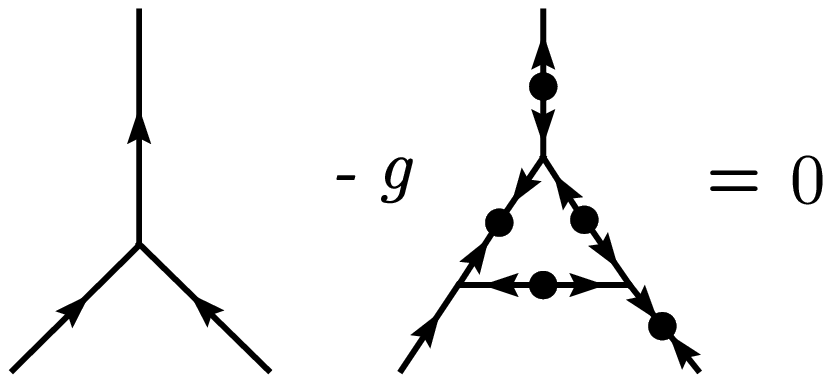,scale=.9}{The equation of motion \eq{sphereeom}. 
The blobs with arrows in the inverse directions
to Fig.\,\ref{fig3} represent $g_{ab}$, the inverse of $g^{ab}$.
\label{fig6}}
An action which leads to the equations of motion \eq{sphereeom} can be constructed, for example, 
by summing the squares of the equations of motion with appropriate contractions of the indices
by $g_{ab},\,g^{ab}$. 
Using the identity \eq{threethreej}, it can be shown that \eq{coftwosphere} is actually a solution 
to \eq{sphereeom} if
\be
\label{avalue}
A=\frac{1}{\sqrt{g\, \sixj 1LL1LL }}=\sqrt{\frac{L(L+1)(2L+1)}{g(L^2+L-1)}}.
\ee

\section{A scalar field theory on a fuzzy two-sphere}
\label{fuzzytwosphere}
In the continuum case, the spherical harmonic functions give the independent functions on a two-sphere.
They are labeled with the SO(3) indices $(j,m)$, where the SO(3) is 
the rotational symmetry in the three-dimensional Euclidean space embedding the two-sphere.
In the discussions in Section\,\ref{themodel}, $f_a$ denotes the functions on a fuzzy space. 
Therefore it would seem reasonable to 
consider the solutions \eq{assofc}, \eq{seriessol} to represent the fuzzy two-sphere,
because the ranges of the indices match between the spherical harmonics and the 
model. But I have not succeeded in this direction.  The main reason for this failure is that it is 
difficult to obtain the spectra of the Laplacian $-j(j+1)$ from the solutions \eq{assofc}, \eq{seriessol}. 
For example, using the properties of the $3j$- and $6j$-symbols,  
one finds
\bea
\label{evalcc}
C^0_{akl}C^0_{bl'k'}g_0^{kk'}g_0^{ll'}=\frac{1}g g^0_{ab},
\eea
where the spins of $a,b \leq 2L$.
There is no dependence on $j$ in \eq{evalcc}. I have computed some other invariants, but have not found the appropriate dependence on $j$.

A way to construct successfully a scalar field theory on a fuzzy two-sphere can be obtained from
the solutions \eq{coftwosphere}, \eq{avalue}. 
This construction is almost similar to the original one \cite{Madore:1991bw}. 
In the paper \cite{Madore:1991bw}, the kinetic term of a scalar field theory is obtained from 
the quadratic Casimir.
The generator $J_i$ of SO(3) in the spin $L$ representation has two spin $L$ indices and a spin 1. 
This index structure is the same as ${C^0_{(1,m_1)\,(L,m_2)}}^{(L,m_3)}$, and the kinetic term
can be constructed in a similar way. 
An extra issue in the present model is that the indices contain $(1,m)$ as well as $(L,m)$.
It must be checked that the scalar field components with these extra indices 
do not destroy the wanted spectra.

Let me consider the following four-index invariant,
\be
\label{defofk}
K_{b'a}^{a'b}=-{C^0_{ia}}^{a'} {C^0_{i'b'}}^b g_0^{ii'},
\ee
shown in Fig.\,\ref{fig7}. 
\EPSFIGURE{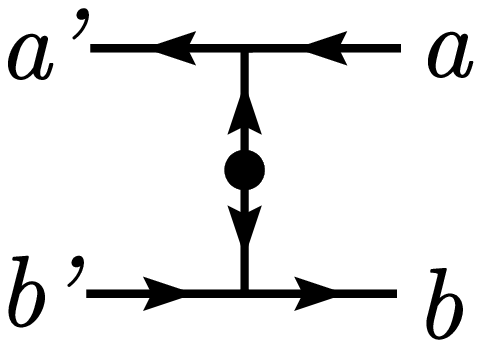,scale=.7}{The quantity  $K_{b'a}^{a'b}$ \eq{defofk}.
\label{fig7}}
This invariant can be regarded as an operator on $\phi^a_b$, 
\be
K_{b'a}^{a'b} \phi^a_b,
\ee
where $\phi^a_b$ will be identified as a scalar field on a fuzzy two-sphere.

Since the only non-vanishing components of ${C^0_{ab}}^c$ are \eq{coftwosphere}, 
the operator $K_{b'a}^{a'b}$ is non-trivial only on $\phi^{(L,m_1)}_{(L,m_2)}$, while it vanishes on 
$\phi^{(L,m_1)}_{(1,m_2)}$,  $\phi^{(1,m_1)}_{(L,m_2)}$,  $\phi^{(1,m_1)}_{(1,m_2)}$.
From the composition rule of two spins, 
the scalar field $\phi^{(L,m_1)}_{(L,m_2)}$ can be decomposed into the components with
total spins $J=0,1,\cdots,2L$.
The scalar field with a total spin $J$ has the components, 
\be
(\phi_{J,m})^{(L,m_1)}_{(L,m_2)}= g_0^{(L,m_1)\, (L,-m_1)} \threej LLJ{-m_1}{m_2}{m}.
\ee
By using the following identities of the $3j$- and $6j$-symbols,
\bea
\sum_{h,m_h} (-1)^{m_h}(2h+1)\sixj{j_1}{J_1}h{J_2}{j_2}f \threej{j_1}{J_1}h{m_1}{M_1}{-m_h} 
\threej{j_2}{J_2}h{m_2}{M_2}{m_h} \cr
=(-1)^{j_2+J_1}\sum_{m_f} (-1)^{m_f} \threej{j_1}{j_2}f{m_1}{m_2}{-m_f}
\threej{J_1}{J_2}f{M_1}{M_2}{m_f},
\eea
and 
\be
\sum_{m_1,m_2}\threej{j_1}{j_2}{j_3}{m_1}{m_2}{m_3} \threej{j_1}{j_2}{j'_3}{m_1}{m_2}{m'_3}=\frac1{2j_3+1}
\delta_{j_3j_3'}\delta_{m_3m_3'},
\ee
one can show that $\phi_{J,m}$ is an eigen vector of the operator $K_{b'a}^{a'b}$, 
\be
K_{b'a}^{a'b} (\phi_{J,m})_b^a
=(-1)^{J} A^2 \sixj LLJLL1 (\phi_{J,m})^{a'}_{b'}=\frac{J(J+1)-2L(L+1)}{2g(L^2+L-1)} (\phi_{J,m})^{a'}_{b'}.
\ee
Let me define the kinetic operator $\tilde K$ by
\be
\label{defofktilde}
\tilde K_{b'a}^{a'b}= K_{b'a}^{a'b}+\frac{L(L+1)}{ g (L^2+L-1) }I_{b'a}^{a'b},
\ee    
where $I_{b'a}^{a'b}$ denotes the identity operator, whose graphical representation is shown in Fig.\,\ref{fig8}.
\EPSFIGURE{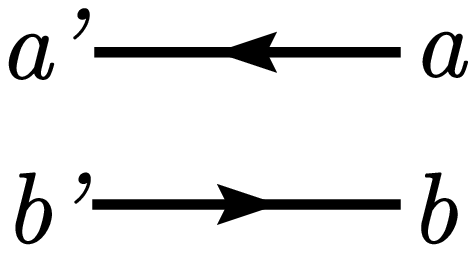,scale=.7}{The identity operator $I_{b'a}^{a'b}$.
\label{fig8}}
The spectra of the operator $\tilde K$ in the $\phi_{(L,m_1)}^{(L,m_2)}$ sector
are given by $J(J+1)/2g(L^2+L-1)\ (J=0,1,\cdots,2L)$. This is the spectra of a massless free scalar field
on a fuzzy two-sphere of a size $\sqrt{2g(L^2+L-1) }$. 
The fields in the other sectors $\phi^{(L,m_1)}_{(1,m_2)}$,  $\phi^{(1,m_1)}_{(L,m_2)}$,  
$\phi^{(1,m_1)}_{(1,m_2)}$ obtain a mass of approximately $\sqrt{1/g\,}$ on account of the identity operator
in \eq{defofktilde}. 
These fields components can be physically decoupled by regarding $g$ to be small enough\footnote{Since the model
has no scales, only a relative scale is physically relevant.}.  

\section{Discussions and comments}
\label{discussions}
In this paper, I studied models with a three-index variable and discussed whether they can be used
as models for dynamical fuzzy spaces. The classical solutions of the models contain many solutions with
Lie group symmetries. 
I considered the solutions with SO(3) symmetry as specific examples, and 
have constructed a scalar field theory on a fuzzy two-sphere. 
The construction in Section\,\ref{so3case} and \ref{fuzzytwosphere} would be straightforwardly 
applied to the general SO($n$) group to obtain the fuzzy complex quadratics discussed in \cite{Dolan:2003th}.

The Lie group symmetries can be different from the orthogonal group, provided that the invariant tensors
can be taken real.  Moreover even in the SO(3) case, the model contains other solutions than 
what were used in the construction of a fuzzy-two sphere. 
It would be obviously interesting to find the interpretation of the other solutions as fuzzy spaces. 

In the construction of a scalar field theory on a fuzzy two-sphere in Section\,\ref{fuzzytwosphere}, 
the scalar field was introduced as an additional degree of freedom, and 
the construction was rather adhoc. 
It is clear that the dynamics of ${C_{ab}}^c$ and $g^{ab}$ is more interesting. One could
analyze their quadratic fluctuations around the classical solutions.
Some of the fluctuations would be identified as scalar fields, and 
higher-spin fields would be also found.
One might suspect that the models contain too many fields,
but this might turn out to be nice, since it was recently argued that higher-spin fields must appear 
in gauge theory on non-associative fuzzy spaces \cite{Ramgoolam:2003cs, deMedeiros:2004wb}.

Actions were not explicitly given, because the main interest of this paper was the classical
solution. On the other hand, 
actions will be needed to perform the analysis in the preceding paragraph and also to study
the quantum properties of the models. 
The quantum process could describe the transitions between distinct fuzzy spaces with different symmetries, 
and is worth to study.
Considering a real action with complex variables would be also interesting, 
since the discussions on real variables in this paper can be essentially applied also to the complex case
and the solutions can be constructed more freely.

The argument about the fuzzy general coordinate transformation has remained inconsistent. 
In the discussions about the fuzzy general coordinate transformation in Section\,\ref{themodel}, 
$f_a$ denote the functions on a fuzzy space. However, 
in the successful construction of the scalar field theory in Section\,\ref{fuzzytwosphere}, 
the scalar field has two indices and cannot be identified with $f_a$.
Another formulation of field theory consistent with the argument or another interpretation of 
the fuzzy general coordinate transformation seems to be required.

It is shown in Section\,\ref{thesolutions} that a lot of solutions can be constructed from the invariant 
tensors of Lie groups. A question is how many of the solutions have the Lie group symmetries.  
If most of the solutions do, it would be interesting to use the models as fuzzy higher dimensions 
\cite{Aschieri:2003vy}-\cite{Aschieri:2005wm} to explain the origin of the symmetries in our world.
Incorporation of fermionic degrees of freedom and supersymmetry will be also interesting as phenomenology.

\acknowledgments{The author was supported by the Grant-in-Aid for Scientific Research No.13135213 and No.16540244
from the Ministry of Education, Science, Sports and Culture of Japan.}

\end{document}